# Single Acquisition Label-free Histology-like Imaging with Dual Contrast Photoacoustic Remote Sensing Microscopy

**Benjamin Ecclestone,**[a] **Deepak Dinakaran**[b,c]**, Parsin Haji Reza,**[a*]

[a] University of Waterloo, Faculty of Engineering, Systems Design Engineering, 200 University Ave W, Waterloo, ON N2L 3G1, Canada
[b] Department of Oncology, University of Alberta, 8440 112 St. NW, Edmonton, T6G 2R7, Alberta, Canada
[d] Illumisonics, 22 King Street S Suite 300, Waterloo, N2J 1N8, Canada

**Abstract**

**Significance:** Histopathological analysis of tissues is an essential tool for grading, staging, diagnosing, and resecting cancers and other malignancies. Current histopathological imaging techniques require substantial sample processing, prior to staining with hematoxylin and eosin (H&E) dyes, to highlight nuclear and cellular morphology. Sample preparation and staining is resource-intensive and introduces potential for variability during sample preparation.

**Aim:** We present a novel method for direct label-free histopathological assessment of formalin fixed paraffin embedded tissue blocks, and thin tissue sections using a dual contrast Photoacoustic remote sensing (PARS) microscopy system.

**Approach:** To emulate the nuclear and cellular contrast of H&E staining, we leverage unique properties of the PARS system. Here, the ultraviolet excitation PARS microscope takes advantage of DNA's unique optical absorption to provide nuclear contrast analogous to hematoxylin staining of cell nuclei. Concurrently, the optical scattering contrast of the PARS detection system is leveraged to provide bulk tissue contrast reminiscent of eosin staining of cell membranes.

**Results:** We demonstrate the efficacy of this technique by imaging human breast tissue and human skin tissues in formalin fixed paraffin embedded tissue blocks and frozen sections respectively. Salient nuclear and extra-nuclear features are captured such as cancerous cells, glands and ducts, adipocytes, and stromal structures such as collagen.

**Conclusions:** The presented dual contrast PARS microscope enables label-free visualization of tissues with contrast and quality analogous to the current gold standard for histopathological analysis. Thus, the proposed system is well positioned to augment existing histopathological workflows, providing histological imaging directly on unstained tissue blocks and sections.

*Parsin Haji Reza**, E-mail: phajireza@uwaterloo.ca

Histopathological analysis of tissues is an essential tool in the diagnoses and treatment of cancers and other malignancies. Histopathology is performed intraoperatively or for diagnostics, to provide visualizations of nuclear structures and microscopic tissue morphology. This enables identification of tissue type, differentiation of pathological or cancerous from healthy tissue, and necrotic from living tissues. Intraoperatively, histopathology helps to guide surgeons in excising cancerous regions while removing only a minimal margin of healthy tissue, thereby preserving the organ's function to the greatest extent [1]. Outside the operating room, histopathological analysis aids in decisions on adjunct treatments, and diagnosing, grading and staging malignancies. The gold standard for performing histopathological imaging is formalin fixed paraffin embedded (FFPE) preparation, followed by hematoxylin and eosin (H&E) staining [2]. Although, frozen preparation methods have become a more common technique in recent years [3]. Though the exact processing differs significantly, the basis of the frozen and paraffin methods are equivalent. First, excised surgical tissues or biopsies are deposited into a substrate forming a solid tissue block [2,3]. Once embedded, thin sections of tissue are shaved from the block using a microtome. These thin sections are then fixed to a microscope slide and finally stained with immunohistochemical dye, usually H&E. Specifically, hematoxylin dye targets nuclei, while eosin targets cell cytoplasm, collagen, and a few other extracellular structures [2]. While this technique provides excellent visualizations compatible with conventional transmission mode light microscopy, sectioning and staining tissues is an involved and resource intensive process. Producing thin sections from tissue blocks requires high precision, sometimes necessitating multiple attempts to recover an adequate slice [3]. Once collected, sections must be fixed to a slide and stained, a process which is subject to some variability, potentially contributing to imaging artifacts [3,4]. This time-consuming procedure can delay diagnosis by days or even weeks in some instances [5]. In the case of frozen preparations, staining and sectioning accounts for over one third of the total preparation time [6]. These process limitations lend towards non-ideal clinical practices and poorer patient outcomes.

In the ideal case, contrast similar to that provided by H&E staining could be acquired label-free directly from unstained tissue preparations. By extension, such a technique would not modify samples during imaging, preserving tissues for further immunohistochemical analysis. For maximal utility, this ideal system should not be limited to a single sample format, but should perform on a variety of preparations, such as FFPE tissue blocks, FFPE tissue sections, and frozen sections. Visualizing tissue morphology directly on preserved samples would mitigate sectioning and staining requirements saving time and resources. Thus far, several pragmatic hurdles have prevented direct histology-like imaging of unstained FFPE and frozen preparations. While thin sections are translucent and therefore compatible with transmission mode imaging, FFPE blocks are typically a few millimeters thick and opaque. Thus, the sample morphology necessitates a reflection mode modality. Moreover, capturing high fidelity nuclear and cellular features, without exogenous contrast agents represents a substantial hurdle. Most popular techniques such as fluorescence microscopy [7], microscopy with ultraviolet surface excitation [8], and light sheet microscopy [9] require exogenous dyes. Despite these challenges a few modalities have shown histology-like imaging capabilities in tissue preparations. Stimulated Raman scattering (SRS) has produced promising label-free histological-like results [10]. However, this has not been shown in FFPE tissue blocks. Conversely, Optical coherence tomography (OCT) has shown label-free imaging capabilities in a variety of sample formats [11]. However, the optical scattering contrast of OCT does not readily provide chromophore specific visualization similar to H&E processing [11]. One alternative technique, optical resolution photoacoustic imaging provides highly chromophore specific contrast [12]. Recently, it has been applied to histological imaging in a variety of preserved samples [12]. This technique leverages the photoacoustic effect to capture the intrinsic optical absorption contrast of tissues. In photoacoustic imaging, a pulsed excitation laser is used to deposit localized optical energy into a sample, which then undergoes thermo-elastic expansion [12]. The localized thermoelastic expansion results in an outward propagating acoustic pressure wave, which is traditionally detected with an acoustically coupled ultrasound transducer [12]. Unfortunately, the requirement for acoustic detection presents several challenges. Acoustic transducers are typically bulky, have known inter-operator technique reliability issues, and sometimes require immersion in a coupling media such as water to function [12].

Recently, Photoacoustic Remote Sensing (PARS) microscopy has emerged as an all-optical non-contact alternative to traditional optical resolution photoacoustic microscopy. PARS replaces the acoustically coupled ultrasound transducer with a detection laser [13]. Photoacoustic signals are then detected as pressure induced modulations in the backscattered magnitude of the detection beam [13]. Observing backscattering in a reflection mode architecture lends PARS to imaging thick samples [13]. Moreover, PARS may provide chromophore specific contrast by selecting excitation wavelengths to target unique biomolecule absorption spectra [14]. Applied to histological imaging, PARS has successfully used ultraviolet (UV) excitation to capture nuclear structures [15,16,17]. This technique is efficacious in providing hematoxylin like contrast in frozen sections, and paraffin embedded blocks and sections [15,16,17]. However, these UV-PARS visualizations lack some essential diagnostic details. In order to provide the full picture, bulk tissue contrast analogous to eosin staining of cell membranes is required to provide structure and reference for the visualized nuclei. This is of particular relevance in tissues with low nuclear density, such as adipose, where the adipocytes are made up of sparse nuclei surrounded by largely lipid-containing vacuoles in the cytoplasm. Previously, PARS has provided complete H&E emulation by using a tunable excitation source to independently target the absorption peaks of DNA and cell membrane structures [16,17]. While effective in both thin sections and tissue blocks, this technique was largely limited in field of view, resolution, and imaging speed, since it required the use of a slow (1Khz) tunable excitation source [16,17]. In this manuscript, we propose to leverage the optical scattering microscope architecture of the PARS detection system to provide bulk tissue contrast, while using an ultraviolet excitation to capture nuclear structures. For the first time, we use the unique ability of PARS microscopy to concurrently provide both label-free absorption and scattering contrast simultaneously. To the extent of our knowledge, PARS is the only optical imaging technique able to provide dual contrast optical scattering and optical absorption simultaneously. We capture eosin-like visualizations by imaging the disparity in the optical scattering of the tissue substrate and embedded tumor tissue. Concurrently, the ultraviolet excitation is used to highlight nuclear morphology, analogous to hematoxylin staining. Combining the PARS optical scattering and absorption images, we provide emulated H&E like visualizations in a single acquisition. With the proposed technique, images are captured approximately 100 times faster than equivalent visualizations with the previously reported multiwavelength system. We demonstrate the proposed method in thin human tissue sections, and FFPE blocks. Employed in a clinical setting, the proposed technique would allow H&E like images to be captured label-free in embedded tissue samples. Thus, PARS is well positioned to supplement existing tissue analysis techniques, potentially mitigating the requirement for sectioning and staining tissues prior to analysis.

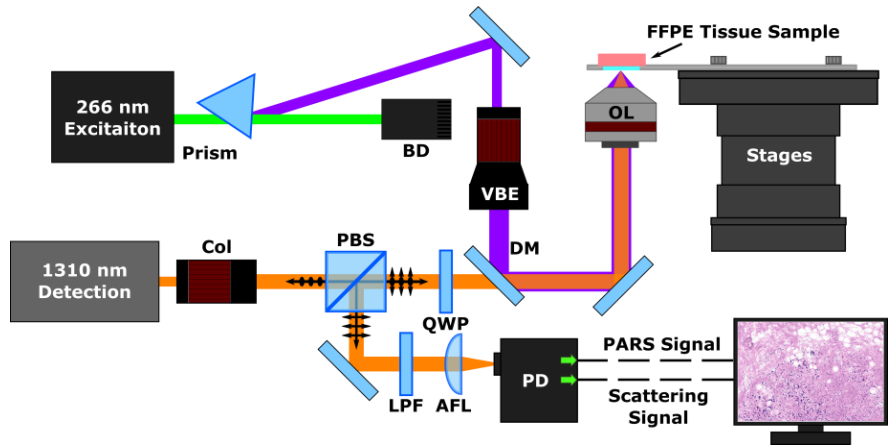

**Fig. 1** Simplified experimental system setup. Component labels are as follows: dichroic mirror (DM), variable beam expander (VBE), beam dump (BD), objective lens (OL), collimator (Col), quarter waveplate (QWP), polarized beam splitter (PBS), long-pass filter (LPF), aspheric focal lens (AFL), photodiode (PD), Spatial Filter (SF).

The architecture of the experimental system is illustrated in Fig. 1. The UV PARS excitation is provided by a 266 nm, 50 kHz, 500 ps pulsed laser (WEDGE XF 266, Bright Solutions). The 266 nm excitation beam is first separated from residual 532 nm output, using a prism (PS862, ThorLabs Inc.). Once isolated, the excitation beam is expanded (BE05-266, ThorLabs Inc.) and combined with the detection via dichroic mirror (HBSY234, ThorLabs Inc.). The PARS detection and optical scattering source is a 1310 nm continuous-wave superluminescent diode (S5FC1018P, ThorLabs Inc). Collimated detection light is passed through a polarizing beam splitter and quarter wave plate, then combined with the excitation. Once combined, the beams are co-focused onto the sample with a 0.5 NA reflective objective (LMM-15X-UVV, ThorLabs Inc.). Back scattered detection light returns to the quarter wave plate by the same path as forwards propagation. Upon return, the polarizing beam splitter redirects the detection beam through spatial and chromatic filters towards the photodiode. In this architecture, two separate outputs are collected from the photodetector. The first is an unamplified and unfiltered output, corresponding to the PARS scattering contrast attributed to the 1310 nm detection source. The second is a band-pass filtered and amplified signal, which isolates nanosecond scale modulation in the optical scattering. This signal is the characteristic PARS absorption contrast. The two signals are captured simultaneously using a multi-channel high-speed digitizer (RZE-004-300, Gage Applied). To form a complete frame, the mechanical stages are used to raster scan the sample across the imaging head. When scanning, the excitation is pulsed at 50 kHz, while the velocity of the stages is modified to achieve the desired lateral spacing between interrogation points. At each excitation pulse, the PARS scattering and absorption amplitudes, and location are recorded. To reconstruct the PARS absorption and scattering images, the raw data is imposed onto a Cartesian grid based on the location signal at each interrogation. Color mapping analogous to hematoxylin and eosin staining is then applied to the PARS absorption and scattering images respectively. To form the final emulated H&E image, a basic linear mixing is performed combining the contrast of the absorption and scattering frames.

We first apply the proposed technique to imaging unstained thin tissue sections (acquired under Research Ethics Board of Alberta (HREBA.CC-18-0277) and University of Waterloo Health Research Ethics Committee (Humans 40275), patient consent was waived samples were archival tissue not required for diagnostic purposes, and no patient identifiers were provided to the researchers.). Here, human skin tissues provide a one-to-one comparison between the proposed technique, UV PARS imaging and brightfield H&E preparation (Fig. 2). These samples were previously used to provide the first direct comparison between PARS and H&E imaging [18]. As outlined in the corresponding publication, unstained frozen skin tissue sections were collected and scanned with the PARS microscope [18]. Following PARS imaging, the unstained sections were stained and imaged with a standard light microscope [18]. The corresponding H&E and PARS visualizations of a small region of this tissue are seen in Fig. 2 (a) and Fig. 2 (b) respectively. This high-resolution close-up of a tissue sample with basal cell carcinoma shows a characteristic invasive front of the tumor, made up of unorganized sheets of cells with aberrant, atypical nuclei. This particular field-of-view captures the clinically relevant boundary between healthy tissue and dense tumor tissue, which is otherwise known as the tumor margin. Visualizing this margin allows clinicians to determine if complete resection of the tumor is achieved. The PARS image accurately captures the nuclear structure in this region Fig. 2 (b), providing a one-to-one match between individual nuclei within the H&E image Fig. 2 (a). While this contrast successfully highlights features including the tumor margin, and localized nuclear atypia, the result is equivalent to using only hematoxylin stain, and thus is not as informative as traditional H&E. Therefore, in contrast to the nuclear dense tumor, the non-nuclear morphology of the healthy tissue in the upper left of the image is poorly represented (Fig. 2 (b)). To remedy this, we apply the proposed technique to obtain eosin-like visualizations. We mix the PARS scattering contrast of the 1310 nm detection with the PARS UV absorption image. The resulting emulated PARS H&E image is shown in Fig. 2 (c). In contrast to the pure PARS UV absorption results, the emulated H&E visualization is nearly identical to that of the stained slide. Here, the extranuclear structures of the nuclear sparse tissue in the upper left of the image are clearly visible. In these nuclear-sparse areas, the PARS scattering contrast allows connective tissue stroma to be distinguished from voids of likely adipose tissue. With the full context of the H&E simulated staining, this slide can be interpreted as the tumor margin (bottom right) invading into regions of subcutaneous tissue. Applied in a clinical setting, directly imaging preserved tissues with this technique, could enable such interpretations without the need for staining.

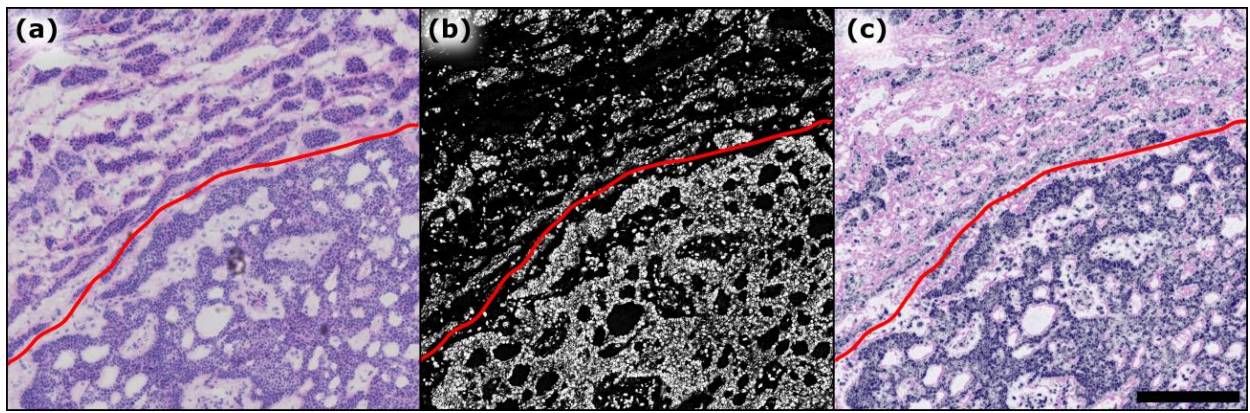

**Fig. 2** One-to-one comparison of bright-field hematoxylin and eosin (H&E) imaging, direct PARS and the proposed PARS absorption and scattering technique in human skin tissue with basal cell carcinoma (BCC). **(a)** Bright field image of H&E stained tissue with BCC demonstrating the border of invasive cancer (bottom of red border) versus normal tissue (top of red border). **(b)** PARS image of the same unstained sample, red border denotes the cancer boundary. **(c)** Combined PARS absorption and scattering image of the same unstained sample, red border denotes the cancer boundary. Scale Bar: 200 µm.

Applied in FFPE tissue blocks, this technique successfully recovers both nuclear and bulk tissue contrast Fig. 3. The PARS image capturing nuclear structures with hematoxylin like contrast is shown in Fig. 3 (a). In isolation, this provides clinically relevant features including inter- nuclear density, cross-section, morphology and internuclear spacing. However, there is a distinct lack of extranuclear tissue morphology. To provide the missing structural details, we therefore leverage the optical scattering contrast of the PARS detection (Fig. 3 (b)). Based on the difference in scattering properties between the sample and the paraffin substrate, we capture eosin like visualizations within the FFPE tissue block. In this case, the scattering image reveals a large region of adipose tissue in the upper right, and a denser region of tissue in the lower left. While these details are of some clinical relevance, without the underlying nuclear structures, the regions of tissue are non-diagnostic. Therefore, we provide a fully emulated H&E image with nuclear and extra nuclear contrast. To do so, the PARS scattering, and absorption images are combined using a basic linear mixing technique. The resulting fully emulated H&E image is shown in (Fig. 3 (c)) providing combined nuclear and bulk tissue structures. With full context of the H&E simulated staining, we can identify further clinical details such as the connective tissue stroma, made up mainly of eosin-staining collagen fibers. The connective tissue stroma can also now be distinguished from interspersed vacuoles, which in FFPE samples such as this most often represent adipose tissue. Observing smaller regions of the FFPE breast tissue with the PARS emulated H&E visualization (Fig. 3 (d) and Fig. 3 (e)), the morphology of mammary glands and ducts are clearly visible. These enhanced regions reveal normal postmenopausal breast tissue with sparse and atrophic glands surrounded by connective tissue. With the information gathered about the nuclear (hematoxylin staining) structures and connective tissue (eosin staining), this tissue block is clearly identifiable as a normal post-menopausal breast tissue specimen. We emphasize that this image is captured directly on a FFPE tissue block, not a thin tissue section. This is the first technique to provide single acquisition H&E like visualizations in such FFPE tissue blocks.

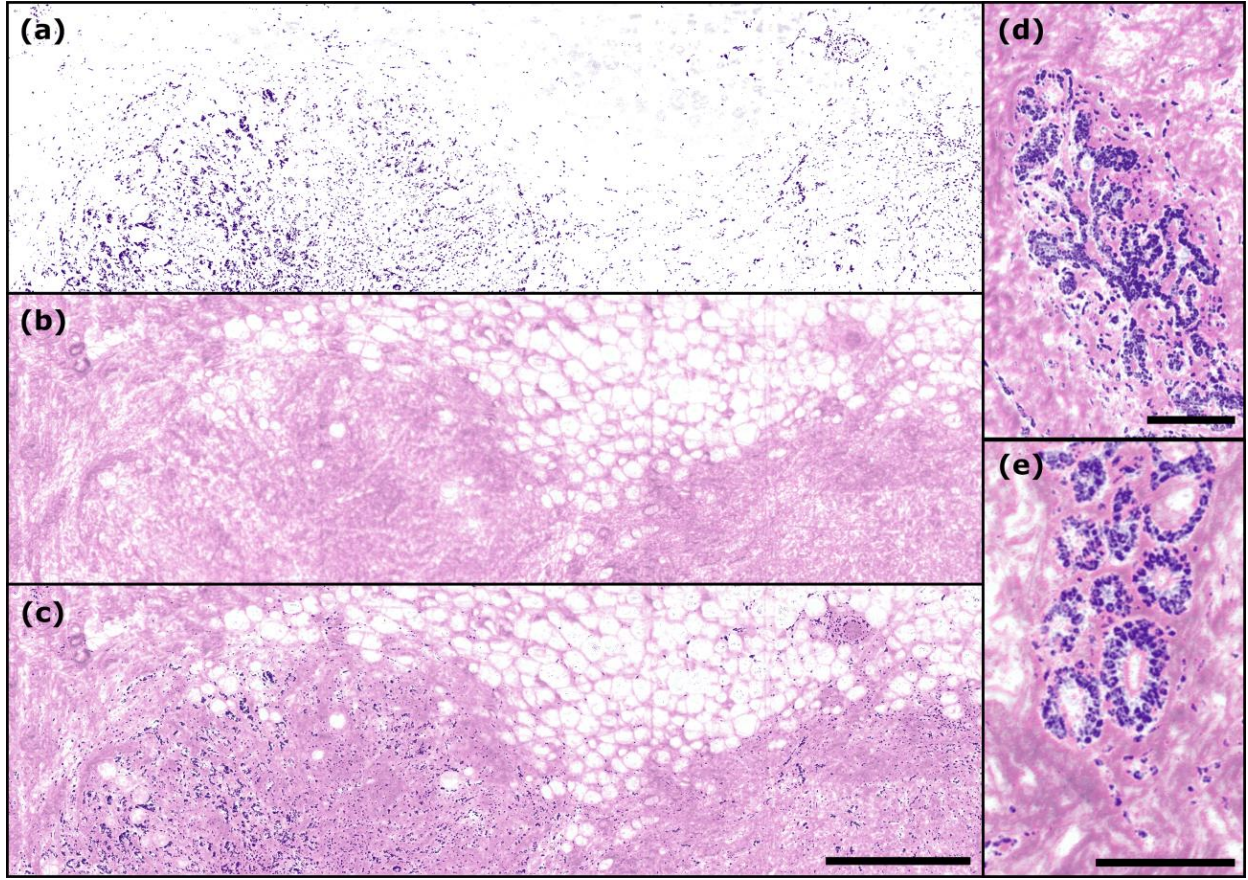

**Fig. 3** Individual and combined PARS absorption and scattering images of FFPE human breast tissues in block form, providing emulated hematoxylin and eosin (H&E) stain contrast. **(a)** UV-PARS image of nuclei providing contrast analogous to hematoxylin staining of cell nuclei. The distribution of nuclei in the breast tissue specimen is evident, but by itself not diagnostic. **(b)** PARS scattering image providing visualizations analogous to eosin staining of cell membranes. This image reveals regions of adipose (clear voids) and dense, collagen rich stromal tissue (pink-colored), and the organization of these structures in relation to each other **(c)** Combined ((a) and (b)) emulated H&E PARS image of normal postmenopausal breast tissue with sparse and atrophic glands with higher nuclear densities (bottom left and right) with mainly adipose (top) and connective tissue (false-colored pink) comprising the majority of the sample. Scale Bar: 500 $\mu m$. **(d-e)** Combined emulated H&E PARS images of normal postmenopausal breast tissue with sparse and atrophic glands surrounded by connective tissue. **(e)** Scale Bar: 100 $\mu m$. **(d)** Scale Bar: 125 $\mu m$.

In comparison to previous PARS based emulated H&E imaging, the proposed method provides several benefits. Previous implementations developed by our group leverage a multi-wavelength tunable excitation to capture hyperspectral images of several chromophores in the tissues [16,17]. In contrast, the proposed technique requires only a single acquisition to be captured as the PARS absorption and scattering images are acquired concurrently. As a result, the scattering and absorption images are perfectly co-registered. Moreover, with bulk tissue contrast captured by the detection source, only UV excitation is required to capture nuclear structures. Thus, a costly and slow tunable excitation laser is no longer required. Instead, one of the many widely available 266 nm sources can be used for the PARS excitation. The primary consideration with this technique then, is the detection source. In this case, the 1310 nm detection is selected to leverage the disparity in the optical scattering properties of the preserved tissue and the surrounding substrate. Based on this difference, we may visualize tissue structures. However, the provided contrast is relatively limited, since deformities in the surface of the samples may also contribute to variations in optical scattering. For similar reasons, the current implementation does not perform well in freshly resected or formalin fixed tissues. In these samples, there is no substrate present to provide a difference in scattering contrast. Moreover, the 1310 nm detection lies within an optical window in tissue [19]. Therefore, moving forwards a shorter visible wavelength detection source may be implemented to provide better scattering contrast in a wider variety of samples. Additionally, moving to a shorter detection wavelength will improve the spatial resolution of the PARS scattering images. Ideally, if a detection wavelength is selected to target high optical scattering in tissues, this method could be applied in freshly resected or formalin fixed samples.

Thus, by isolating and leveraging the different contrast mechanisms provided by the PARS excitation and detection, we provide the first technique for single acquisition recovery of label-free emulated H&E images in preserved tissue blocks and sections. The combination of the two PARS contrast mechanisms provides rapid visualization of critical diagnostic features. While the PARS UV absorption captures nuclear contrast analogous to hematoxylin staining, the PARS scattering captures bulk tissue contrast analogous to eosin staining. Combining the two different images provides a direct emulation of the current gold standard for histopathological imaging. As presented here, these emulated H&E visualizations are provided directly on unstained FFPE blocks, and in thin tissue sections. The images captured with this technique provide comparable contrast and resolution to the gold standard technique. Several salient features such as the nuclear organization, the connective tissue stroma and the relationship between the highly nuclear cells to the connective tissue are important for deriving diagnostic information as the PARS H&E-simulated images highlight here. By visualizing nuclear and bulk tissue structure directly on the thick FFPE specimens, the proposed dual contrast PARS microscope could potentially remove the need for sectioning and staining. This would expedite the histopathological workflow, saving time during the diagnostic process, removing potential for variability between successive tissue block sectioning for thin film slides. Moreover, PARS imaging is non-destructive and does not affect immunohistochemical staining efficacy. As a direct result, tissue blocks, or unstained tissue samples could be imaged with PARS prior to undergoing further immunohistochemical processing, directly reducing premature consumption of small biopsy specimens. Overall, the label-free non-contact microscope presented here holds great potential as an adjunct to existing histopathological workflows. Adopted in a clinical setting this microscope could provide a tabletop device for direct histological assessment of unstained embedded tissues.

*Disclosures*

The authors Benjamin Ecclestone, Deepak Dinakaran and Parsin Haji Reza have financial interest in the illumisonics company which partially supported this work.

*Acknowledgments*

The authors thank the following sources for funding used during this project. Natural Sciences and Engineering Research Council of Canada (DGECR-2019-00143, RGPIN2019-06134); Canada Foundation for Innovation (JELF #38000); Mitacs Accelerate (IT13594); University of Waterloo Startup funds; Centre for Bioengineering and Biotechnology (CBB Seed fund); illumiSonics Inc (SRA #083181); New frontiers in research fund – exploration (NFRFE-2019-01012).

*Code, Data, and Materials Availability*

All data sets and materials relevant to this study is included in this article.

*References*

1. S. Weber et al., "The role of frozen section analysis of margins during breast conservation surgery," *Cancer J Sci Am.* **3**(5), 273-277 (1997).

2. K. Canene-Adams, "Preparation of formalin-fixed paraffin-embedded tissue for immunohistochemistry," *Methods Enzymol.* **533**, 225-33 (2013).

3. H. Jaafar, "Intra-operative frozen section consultation: concepts, applications and limitations," *Malays J Med Sci*. **13**(1), 4-12 (2006).

4. S.A. Taqi et al., "A review of artifacts in histopathology," *J Oral Maxillofac Pathol*. **22**(2), 279 (2018).

5. L. Brown, "Improving histopathology turnaround time: a process management approach," *Curr. Diagn. Pathol.* **10**(6), 444-452 (2004).

6. D.A. Novis and R.J. Zarbo, "Interinstitutional comparison of frozen section turnaround time. A College of American Pathologists Q-Probes study of 32868 frozen sections in 700 hospitals," *Arch Pathol Lab Med*. **121**(6),559-67 (1997).

7. L. Cahill et al., "Rapid virtual hematoxylin and eosin histology of breast tissue specimens using a compact fluorescence nonlinear microscope," *Lab Invest* **98**, 150–160 (2018).

8. F. Fereidouni et al.*,* "Microscopy with ultraviolet surface excitation for rapid slide-free histology," *Nat Biomed Eng* **1**, 957–966 (2017).

9. A. Glaser et al. "Light-sheet microscopy for slide-free non-destructive pathology of large clinical specimens," *Nat Biomed Eng* **1**, 0084 (2017).

10. M. Ji et al. "Rapid, label-free detection of brain tumors with stimulated Raman scattering microscopy," *Sci Transl Med*. **5**(201), 201ra119 (2013).

11. E. Min et al.*,* "Serial optical coherence microscopy for label-free volumetric histopathology," *Sci Rep* **10,** 6711 (2020).

12. G. Stasi and E.M. Ruoti "A critical evaluation in the delivery of the ultrasound practice: the point of view of the radiologist," *Ital J Med.* **9**(1), (2015).

13. T.T.W. Wong et al., "Fast label-free multilayered histology-like imaging of human breast cancer by photoacoustic microscopy," *Sci Adv* **3**(5), e1602168 (2017).

14. P. Haji Reza et al., "Non-interferometric photoacoustic remote sensing microscopy," *Light Sci Appl* **6**, e16278 (2017).

15. S. Abbasi et al., "Chromophore selective multi-wavelength photoacoustic remote sensing of unstained human tissues," *Biomed. Opt. Express* **10**, 5461-5469 (2019).

16. N. J. M. Haven et al., "Ultraviolet photoacoustic remote sensing microscopy," *Opt. Lett.* **44,** 3586-3589 (2019).

17. K. Bell et al.*,* "Reflection-mode virtual histology using photoacoustic remote sensing microscopy," *Sci Rep* **10,** 19121 (2020).

18. B.R. Ecclestone et al.*,* "Improving maximal safe brain tumor resection with photoacoustic remote sensing microscopy," *Sci Rep* **10,** 17211 (2020).

19. B. R. Ecclestone et al., "Histopathology for Mohs micrographic surgery with photoacoustic remote sensing microscopy," Biomed. Opt. Express **12**, 654-665 (2021).

20. H. Zhang et al., "Penetration depth of photons in biological tissues from hyperspectral imaging in shortwave infrared in transmission and reflection geometries," J. Biomed. Opt. **21**(12) 126006 (2016).